%% file: main.tex
\documentclass[runningheads]{llncs}

\usepackage[utf8]{inputenc}
\usepackage{multirow}
\usepackage{hyperref}
\usepackage{dirtytalk}
\usepackage{graphicx}
\usepackage{xcolor}
\usepackage{cite}
\usepackage[normalem]{ulem}

\newcommand{\comment}[1]{}

 \newcommand{\revision}[1]{{#1}}

\newcommand{\reviewer}[1]{}

\begin{document}

\title{
    Performance and Usability of Visual and Verbal Verification of Word-based Key Fingerprints%
    \thanks{This is an accepted manuscript to appear in the proceedings of the 15th International Symposium on Human Aspects of Information Security \& Assurance, HAISA 2021.}
}
\titlerunning{Performance and Usability of Visual and Verbal Fingerprints}

% \author{
%     Lee~Livsey\orcidID{0000-0003-4028-5824} \and
%     Helen~Petrie\orcidID{0000-0002-0100-9846} \and 
%     Siamak~F.~Shahandashti\orcidID{0000-0002-5284-6847} \and 
%     Aidan~Fray
%  }

\author{
    Lee Livsey \and
    Helen Petrie \and 
    Siamak F. Shahandashti \and 
    Aidan Fray 
 }

\authorrunning{L. Livsey et al.}

\institute{Department of Computer Science, University of York, York, UK\\
\email{\{lwl501, helen.petrie, siamak.shahandashti\}@york.ac.uk}}

\maketitle   % typeset the header of the contribution

\begin{abstract}
% %The abstract should briefly summarize the contents of the paper in 15--250 words.
The security of messaging applications against person-in-the-middle attacks relies on the authenticity of the exchanged keys. 
For users unable to meet in person, a manual key fingerprint verification is necessary to ascertain key authenticity. 
Such fingerprints can be exchanged visually or verbally, and it is not clear in which condition users perform best. 
This paper reports the results of a $62$-participant study that investigated differences in performance and perceived usability of visual and verbal comparisons of word-based key fingerprints, and the influence of the individual's cognitive learning style. 
The results show visual comparisons to be more effective against non-security critical errors and are perceived to provide increased confidence, yet participants perceive verbal comparisons to be easier and require less mental effort. 
Besides, limited evidence was found on the influence of the individual's learning style on their performance. 

    \keywords{
        Key Fingerprint Verification \and 
        Verbal and Visual Comparisons \and 
        Usability Evaluation \and 
        Index of Learning Styles (ILS)
    }
\end{abstract}

\input{sections/introduction}
\input{sections/background}

\input{sections/method}

\input{sections/results}

\input{sections/discussion_conclusions}

\bibliographystyle{abbrv}
\bibliography{refs}

% \begin{appendix}
% \input{sections/appendices}
% \end{appendix}

\end{document}

%% file: sections/introduction.tex
\section{Introduction}

The use of secure messaging applications has grown rapidly over the last decade, as users seek to reclaim their privacy.
An as yet unsolved problem, particularly when users are unable to meet in person, is a usable protocol for authenticated key exchange that eliminates the risk of person-in-the-middle attacks.

Current solutions begin with the exchange of a key-dependent verification message via an out-of-band channel (OOB), which assures the integrity of `short' messages~\cite{Kainda2009-if}.
If users can meet in person, they may create an OOB channel %\revision{directly} 
between their devices 
\revision{and %Using the OOB channel, the users' devices can 
automatically verify the authenticity of each other's public key material (e.g. through NFC or scanning a QR code).} This solves the problem %in this special case,
\revision{for the in-person context,} yet \revision{such applications are mainly %meant
intended for remote communication as} it is not always feasible for users to meet in person.% especially since

In the remote setting, the OOB channel cannot be directly implemented between devices. %since the communication channel between devices is not necessarily secure. No guarantee of message integrity at first meeting/ without previous sharing of key material. 
\revision{The solution is to directly involve users in the comparison of their \emph{key fingerprints}}, short strings usually computed through cryptographic hashing of key materials. If the received fingerprint
from the manual OOB channel is identical to that from the communication channel, both users can be assured of the authenticity of the keys they hold and hence the security of their communication. 
% Practical examples of user-facilitated OOB channels include phone calls and publishing the fingerprint on a trusted website, e.g. \url{www.theguardian.com/pgp}. 
Fingerprints are usually encoded into easy-to-use formats such as chunked numbers (e.g. in Signal/WhatsApp), or dictionary words (e.g. in Pretty Easy Privacy (PEP, \url{www.pep.security})) for \revision{Pretty Good Privacy (PGP)} keys. 

Though comparison of fingerprints avoids the requirement to meet in person, it introduces significant potential for human error and opens an additional attack vector for adversaries. The adversary need only identify a near-collision fingerprint with sufficient similarity to the authentic fingerprint that it is likely to be accepted by an inattentive user. This is a considerably easier task than finding full collisions necessary for a successful attack in the in-person setting.

%This paper investigates differences in the performance and perception of usability between visual and verbal comparisons of word-based fingerprints.
%
Historically users tended to compare fingerprints visually, but secure messaging applications increasingly encourage a verbal comparison, a substantially different task that places very different demands upon the user. 
\revision{As there has been no previous investigation of user performance and perceived usability between visual and verbal fingerprint comparisons, a within-participants study 
% with $62$ participants 
is designed to investigate such differences in the context of word-based fingerprints.} 

The study also investigates the influence of an individual's preferred method to receive and process information, known as \emph{cognitive} or \emph{learning style}, as measured by the Visual--Verbal subscale of the Index of Learning Styles (ILS)~\cite{felder1988learning}. 
%\revision{This is also} an important aspect in the 
\revision{It may be that users have a preference for processing information either verbally or visually, which would affect the} development of usable and secure fingerprint verification protocols and to our knowledge is yet to be investigated. 

A within-participants study with 62 participants assessed the effectiveness, efficiency and perceived usability of each comparison mode. The results provide valuable insight and demonstrate a complex picture. The answer of which comparison
mode is best remains unclear, with the more effective comparison mode also perceived to be less usable.

%% file: sections/background.tex
\section{Background and Related Work}\label{background_section}

Usability issues in secure messaging applications have been extensively studied~\cite{Orman2015-mg,Ruoti2015-lg,Ruoti2013-qo,Whitten1999-xd}.
%They were first highlighted by Whitten and Tygar who found that non-specialist users where unable to successfully use PGP 5.0 to send secure emails \cite{Whitten1999-xd}, and similar issues continue to be identified~\cite{Ruoti2013-qo,Orman2015-mg,Ruoti2015-lg}.
Recent work has identified usability issues specific to the authentication \revision{procedures} of modern secure messaging applications. 
Schr{\"o}der et al.\ investigated the usability of Signal and found that from a sample of $28$ computer science students, 21 were unable to successfully verify their recipient's public key~\cite{Schroder2016-mk}. 
Related work identified similar issues with WhatsApp, Viber and Telegram, finding that participants were both unaware of the need to verify their recipient's key and unable to do so without additional instruction~\cite{Herzberg2016-wd,Vaziripour2017-ap}.
  
Dechand et al.\ performed a detailed investigation of textual fingerprint representations, finding that word-based formats led to higher usability scores and increased attack detection rates than the traditional hexadecimal format~\cite{dechand2016empirical}. In a similar study, Tan et al.\ investigated a range of visual and textual fingerprint formats, finding that the performance of visual formats varied and that text-based formats achieved some of the lowest error rates~\cite{tan2017can}.
Both studies simulated visual fingerprint comparisons, with the received fingerprint displayed \revision{on} a business card. 
Though verbal comparisons were mentioned by Dechand et al., neither study performed a comparison between visual and verbal modes.

Studies investigating a range of existing device pairing methods identified interesting differences in usability between visual and verbal fingerprint comparisons, but they involve substantially shorter fingerprints that provide sufficient security only for short-range device pairing scenarios~\cite{Kobsa2009-gw,Kumar2009-fu}. %(whereas in our case fingerprints are substantially longer)

\revision{There has been considerable psychological and educational research into the concept of different cognitive or learning styles, with many different dimensions and models proposed. However, one of the more robust is visual-verbal processing. While the concept of learning style is controversial~\cite{Willingham2015-cg}, and people are undoubtedly flexible in the ways they can process information, they may have preferences which would affect their perception of the usability of an authentication system.}
The Index of Learning Styles (ILS) was developed to gain insight into the preferred learning styles of engineering students and provide recommendations of how teaching can be adapted accordingly~\cite{felder1988learning}. 
The ILS %has been shown to be 
\revision{is} a reliable and valid instrument to assess learning styles, and each of its four dimensions display high test-retest correlation coefficients after intervals of between four weeks and eight months%~\cite{felder2005applications}.
~\cite{Livesay2002-statanalysis,Seery2003-multimodal,Zywno2003-validation,felder2005applications}.
The Visual--Verbal subscale of the ILS assesses individual preference to receive and process information visually (e.g., through pictures and diagrams) or verbally (e.g., through written or spoken-aloud text).

%% file: sections/method.tex
\section{Method}

\subsection{Design}

The study involved a within-participants design with two conditions, with each participant comparing 20 pairs of key fingerprints visually and 20 verbally. 
The order of taking conditions was counterbalanced. Two of the 20 comparisons were simulated attacks and the others were non-attack comparisons. 
A low attack rate was used to avoid raising participants’ awareness of the possibility of attack and because attacks are uncommon in practice. Participants were asked to simulate an authentication task by matching a fingerprint of five words, either visually or verbally. The five words were selected from the Trustwords word base~\cite{pep-ietf-draft-2020}.

Performance was measured by time to make correct comparisons and errors, for both attack and non-attack comparisons. Usability was measured on a set of five-level rating items. Standard usability instruments such as the System Usability Scale (SUS)~\cite{Brooke1996-ko} do not capture all the aspects of the user experience of interest, e.g.\ trust that the comparison provides security and confidence in one’s judgement. Therefore, a specific set of questions was developed (see Table \ref{tab:perceived_usability_dimension}).

\begin{table}
\centering
\caption{Dimensions of perceived usability and related concepts}
\begin{tabular}{|c|l|}
\hline
\textbf{Dimension} & \textbf{Rating items} \\ \hline
\textbf{Efficiency} & 
\begin{tabular}[c]{@{}l@{}}
    I was able to do the comparisons very quickly with this method.\\ 
    Comparisons using this method were unacceptably long.
\end{tabular} \\ \hline
\textbf{Ease of use} & 
\begin{tabular}[c]{@{}l@{}}
    The method was easy to use. \\ 
    The method was unnecessarily complex.
\end{tabular} \\ \hline
\textbf{\begin{tabular}[c]{@{}c@{}}
    Low mental\\ 
    workload
\end{tabular}} & 
\begin{tabular}[c]{@{}l@{}}
    The comparisons did not need much mental effort. \\ 
    I needed to concentrate a lot.
\end{tabular} \\ \hline
\textbf{Confidence} & 
\begin{tabular}[c]{@{}l@{}}
    I would need a lot of technical support to be able to use this method.\\ 
    I am confident that I can make comparisons using this method\\ 
    \quad without making mistakes.
\end{tabular} \\ \hline
\textbf{Repeat use} & 
\begin{tabular}[c]{@{}l@{}}
    Completing the comparisons using this method was annoying. \\ 
    Using this method is worth it for the additional security it provides.
\end{tabular} \\ \hline
\textbf{Trust} & 
\begin{tabular}[c]{@{}l@{}}
    Making comparisons using this method would keep my\\ 
    \quad communications secure. \\ I
    would not trust this method when sending confidential information.
\end{tabular} \\ \hline
\end{tabular}
\label{tab:perceived_usability_dimension}
\end{table}

The Hypotheses investigated were: 
\begin{enumerate}
    \item[$H_{1}$] There is a significant difference in time to make the correct decision between the visual and verbal fingerprint comparisons.
    \item[$H_{2}$] There is a significant difference in the number of errors made
    using the visual and verbal fingerprint comparisons.
    \item[$H_{3}$] There is a significant difference in perceived usability ratings between the visual and verbal fingerprint comparisons. 
    \item[$H_4$] Participants perform significantly better and report significantly greater perceived usability when the comparison mode aligns with their preferred method to receive and process information.
\end{enumerate}

Ethical principles of no harm and informed consent
% and data protection 
were followed and formal ethical approval was obtained from \revision{the authors' departmental ethics committee}.

\subsubsection{Security Assumptions.}\label{security_assumptions}
% Within our study we envisage a targeted user who is security conscious and aware of the threats posed by communication using unauthenticated public keys. Consequently, they will always perform a fingerprint comparison via an OOB channel before first communication with a contact or after notification of a key reset.  We note that such users are unusual %reference to low performance rates of fingerprint comparisons
% but they do exist. Examples include journalists, human rights activists, and businesses who seek to maintain a competitive advantage over their rivals.

The study assumed the adversary randomly generates a large set of public keys before implementing a person-in-the-middle attack. During the attack, they replace the authentic keys with ones from this set that display maximal similarity to the target fingerprint. This study simulated such an adversary using $2^{21.8}$ distinct PGP public keys scraped from PGP key servers, with optimal attacks found to possess fingerprints with three out of five identical words. The structure of the attacks remained consistent throughout, with all differences confined to the third and fourth words,  \revision{which is consistent with previous studies \cite{dechand2016empirical, tan2017can}.} The adversary was also assumed to be unable to manipulate any messages exchanged over the OOB channel. 

\subsection{Participants}

Several methods of participant recruitment were used: through the %[anonymised] 
\revision{University of York}
network, the authors' personal contacts, and through Amazon Mechanical Turk (MTurk).  Participants recruited from local networks were entered into a prize draw, whilst participants from MTurk were paid USD 2.00. Some researchers have raised doubts about the care with which MTurk participants undertake tasks~\cite{Chandler2014-hi}, but others have found that MTurk participants produce data of equal quality to those recruited in more traditional ways~\cite{Thomas2017-mx}.  
Therefore, it was decided to use both more traditional recruitment methods and MTurk and compare data from the two sources. No differences in responses were detected between the two groups (comparisons were made on times, errors and responses to rating questions), so results are presented for the whole sample.

In total, 75 people responded to the study, but data from 13 participants were eliminated: 2 experienced network errors, 8 provided a partial response, and 1 failed to identify a totally mismatching attention check. Data from 2 participants who are dyslexic was also eliminated. Both comparison modes involve reading words, including unusual words, which may be difficult for people with dyslexia. All participants whose data were excluded were still rewarded for their time.
%Such groups present an interesting open problem as their specialised requirements are not fully accounted for within the key verification protocols of secure messaging applications.

\begin{table}[t]
\begin{minipage}{.47\textwidth}
\centering
\caption{Age distribution.}
\begin{tabular}{|c|c|}
\hline
\textbf{Age}      & \textbf{Count} \\ \hline
18--24            & 2              \\ \hline
25--34            & 22             \\ \hline
35--44            & 22             \\ \hline
45--64            & 14              \\ \hline
65 and over       & 1              \\ \hline
Prefer not to say & 1              \\ \hline
\end{tabular}
\label{tab:age_distribution}
\end{minipage}\quad
\begin{minipage}{.47\textwidth}
\centering
\caption{Education background.}
\begin{tabular}{|c|c|}
\hline
\textbf{Highest Education level}              & \textbf{Count} \\ \hline
High School education                         & 9              \\ \hline
Vocational training                           & 4              \\ \hline
Bachelors degree                              & 32             \\ \hline
Postgraduate degree                           & 13             \\ \hline
Other                                         & 3              \\ \hline
Prefer not to say                             & 1              \\ \hline
\end{tabular}
\label{tab:education_background}
\end{minipage}
\end{table}

Data from 62 participants were analysed, 25 men (40\%), 36 women (58\%) and one who identified as non-binary. 
Age ranged from 18--24 to over 65, with the majority being in the 25--44 years range (71\%, see Table \ref{tab:age_distribution}). 
Educational level ranged from high school education to postgraduate degree, with the majority having a bachelors or postgraduate degree (73\%, see Table~\ref{tab:education_background}). 
As the experimental task involves reading and listening, participants were asked whether they had a visual or hearing impairment, none reported any.  For the same reason, participants were asked about their proficiency in English; 98\% (61/62) rated it as good or excellent, and one as average. There were 29 participants recruited via the local networks, all located in the UK except one from the USA. There were 33 participants recruited via MTurk, all in the USA.
Participants responses showed 94\% (58/62) use at least one secure messaging application, and 60\% (37/62) do so every day. Furthermore, 87\% (54/62) of participants agree that `it is important to be able to have private conversations using secure messaging applications', yet 82\% (51/62) of participants have never performed a fingerprint comparison.

\subsection{Materials and Task}\label{materials_task_section} 

A web application was developed to enable participants to interact with mockups of two mobile devices and compare fingerprints, with PEP over PGP used as a template for the secure messaging application.
\revision{PEP was chosen as it includes a word-based fingerprint representation which have} been shown to provide high usability and low error rates. PEP uses a word list called Trustwords to replace every 16 bits of the hashed key with one word from Trustwords, hence resulting in five-word fingerprints to represent 80-bit hashes~\cite{pep-ietf-draft-2020}. PEP is supported by popular email clients such as Mozilla Thunderbird.

\begin{figure}[t]
\begin{minipage}{.47\textwidth}
\centering
\caption{Visual comparison task interface.}
\includegraphics[width=1.0\textwidth]{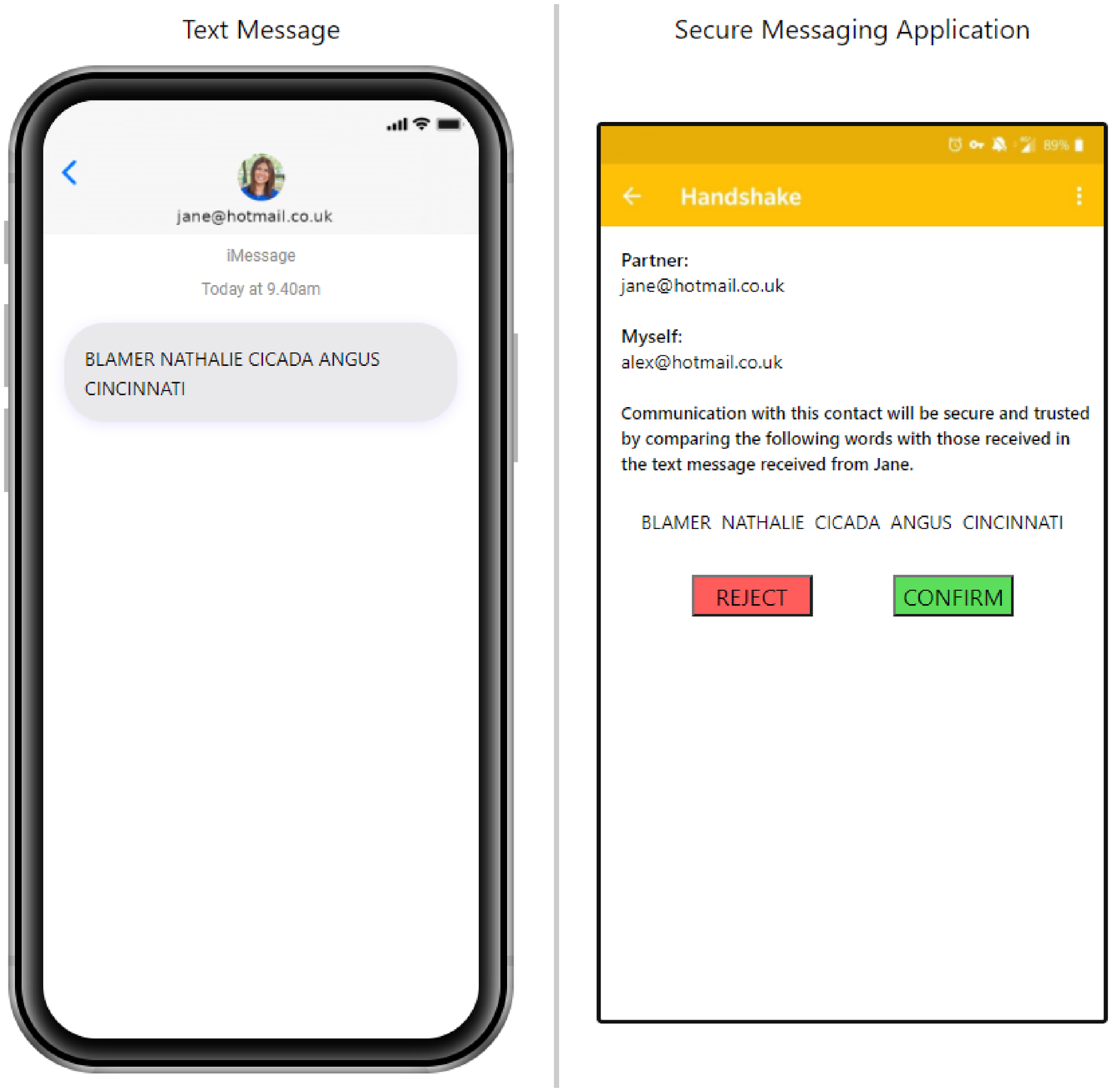}
\label{fig:visual_interface}
\end{minipage}\qquad
\begin{minipage}{.47\textwidth}
\centering
\caption{Verbal comparison task interface.}
\includegraphics[width=1.0\textwidth]{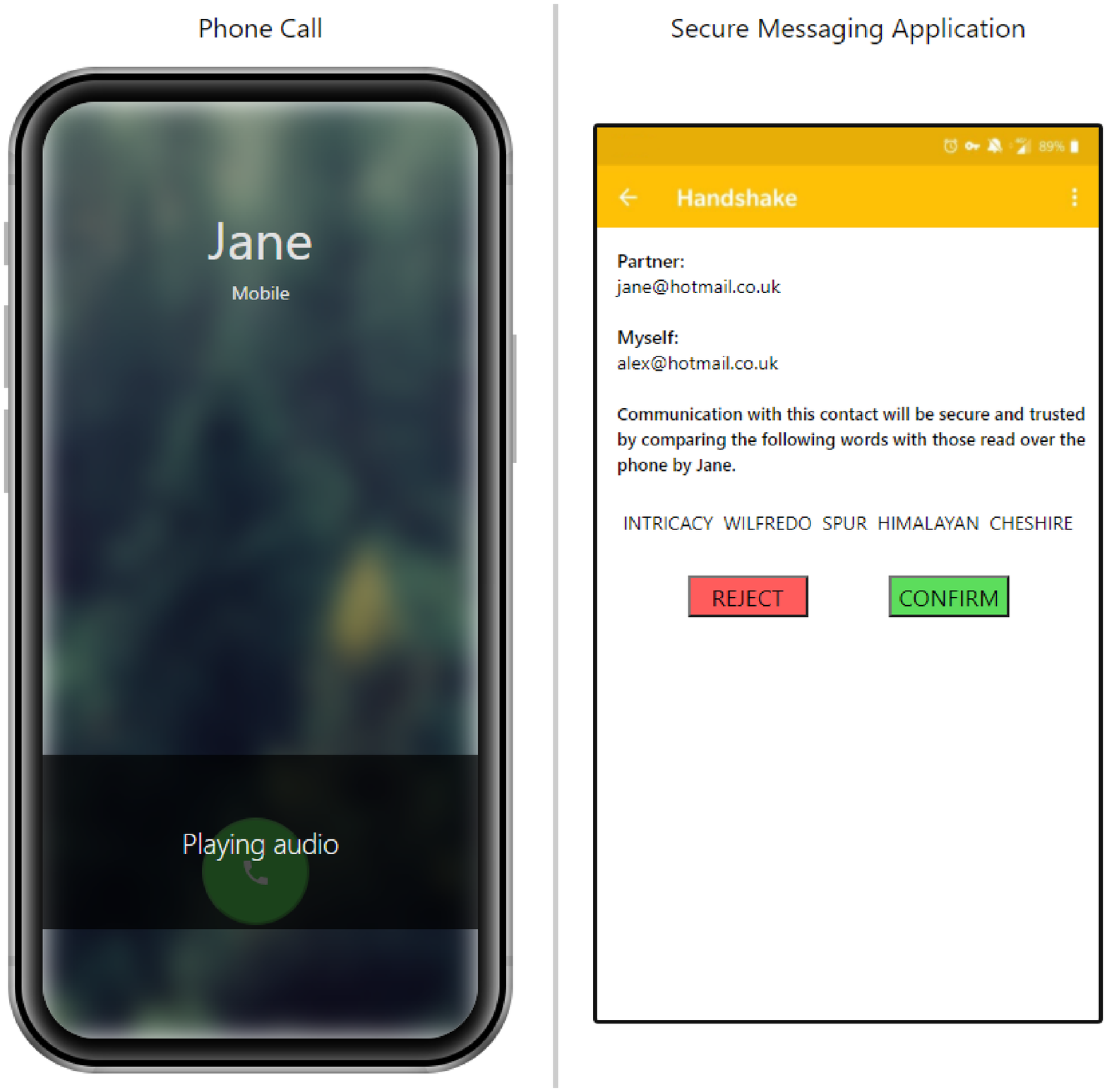}
\label{fig:verbal_interface}
\end{minipage}
\end{figure}

The visual condition simulated a fingerprint exchange by text message (see Fig.~\ref{fig:visual_interface}). The verbal condition simulated an exchange by voice (e.g. by phone) by playing a recorded reading of the words (see Fig.~\ref{fig:verbal_interface}). 
The web application did not allow study completion on small screens, e.g.\ smartphones, that could not display the two virtual devices side by side.
The 11 forced-choice questions of the ILS Visual--Verbal subscale \revision{(see Section \ref{background_section})} were used to measure individuals’ preferences for receiving and processing information. The subscale is scored from -11 (if all questions are answered with a verbal preference) to +11 (if all questions are answered with a visual preference).

A post-task questionnaire assessed the perceived usability of each condition. Six dimensions of usability and related concepts were identified as being of interest and two five-level rating scale items were used to measure each dimension (see Table \ref{tab:perceived_usability_dimension}). The scoring of items was reversed as appropriate so that a high number always indicates high usability.  
A post-study questionnaire asked participants which condition they preferred, their previous experiences using secure messaging applications and also collected demographic information.

\subsection{Procedure}

Before running the main study, a pilot study was conducted with four participants similar in characteristics to the target sample. This led to improvements in the explanation of the task (e.g. to clarify that participants were expected to make multiple comparisons in each condition). Several issues identified in the web application were also resolved. 
The main study procedure was as follows:
\begin{enumerate}
    \item An information sheet explained the aims of the study, described the tasks participants would undertake and the data to be collected. Participants were asked to confirm that they were over 18 and to consent to participation.
    \item Participants were asked two screening questions: if they could view an image displayed upon their device and if they could play and hear a sound clip. This ensured that participants’ devices supported the experimental conditions.
    \item Participants then completed the Visual--Verbal subscale of the ILS.
    \item Participants were randomly assigned to complete either the visual or verbal condition, compared the 20 fingerprints in that condition, and answered a post-task questionnaire to assess the perceived usability of that condition.
    \item The above step was then repeated for the other condition. 
    \item Participants then answered the post-study questionnaire.
    \item Participants were then thanked and provided with the relevant reward.
\end{enumerate}

%% file: sections/results.tex
\section{Results}

Data did not meet the requirements for parametric statistics (normality, homogeneity of variance), so non-parametric statistics were used, with medians and semi-interquartile range (SIQR) as measures of central tendency and spread. To compare between conditions, Wilcoxon related samples non-parametric tests were used. To compare participants with different information styles, Kruskal--Wallis tests were used.

\subsection{Performance: task completion time and errors}
The time to complete correct comparisons did not differ significantly between the visual and verbal modes for either the attack or non-attack trials, as tested by Wilcoxon signed-rank tests for related samples \revision{(see Table~\ref{tbl:times})}. 
Thus $H_{1}$, that there is a difference between the times on the two conditions, was not supported.

\begin{table}
\centering 
\caption{\revision{Median times (seconds) and SIQR on correct comparisons for verbal and visual conditions with Wilcoxon signed rank tests of differences between conditions}}
\label{tbl:times}
\begin{tabular}{|l|c|c|c|c|}
\hline
 & \textbf{Verbal} & \textbf{Visual} & \textbf{Wilcoxon W} & \textbf{p-value} \\ \hline
\textbf{Attack comparisons}     & 5.49 (0.75)   & 5.50 (1.04)  & 0.22            & 0.83                 \\ \hline
\textbf{Non-attack comparisons} & 6.15 (0.55)   & 6.52 (1.96)  & 1.20            & 0.23                 \\ \hline
\end{tabular}
\end{table}

% \begin{figure}[b]
% \begin{minipage}{.47\textwidth}
% \centering
% \caption{\revision{Distribution of mean correct attack comparison times by condition (outliers not shown).}}
% % \includegraphics[width=1.0\textwidth]{./images/plots/efficiency/attack_times.jpeg}
% \includegraphics[width=1.0\textwidth]{./images/plots/efficiency/attack_times.eps}
% \label{fig:attack_times_boxplot}
% \end{minipage}\qquad
% \begin{minipage}{.47\textwidth}
% \centering
% \caption{\revision{Distribution of the mean correct non-attack comparison times by condition (outliers not shown).}}
% % \includegraphics[width=1.0\textwidth]{./images/plots/efficiency/non_attack_times.jpeg}
% \includegraphics[width=1.0\textwidth]{./images/plots/efficiency/non_attack_times.eps}
% \label{fig:non_attack_times_boxplot}
% \end{minipage}
% \end{figure}

\begin{figure}
\begin{minipage}{.47\textwidth}
\centering
\caption{Number of errors \revision{by each participant} on \revision{17} non-attack comparisons}
\includegraphics[width=1.0\textwidth]{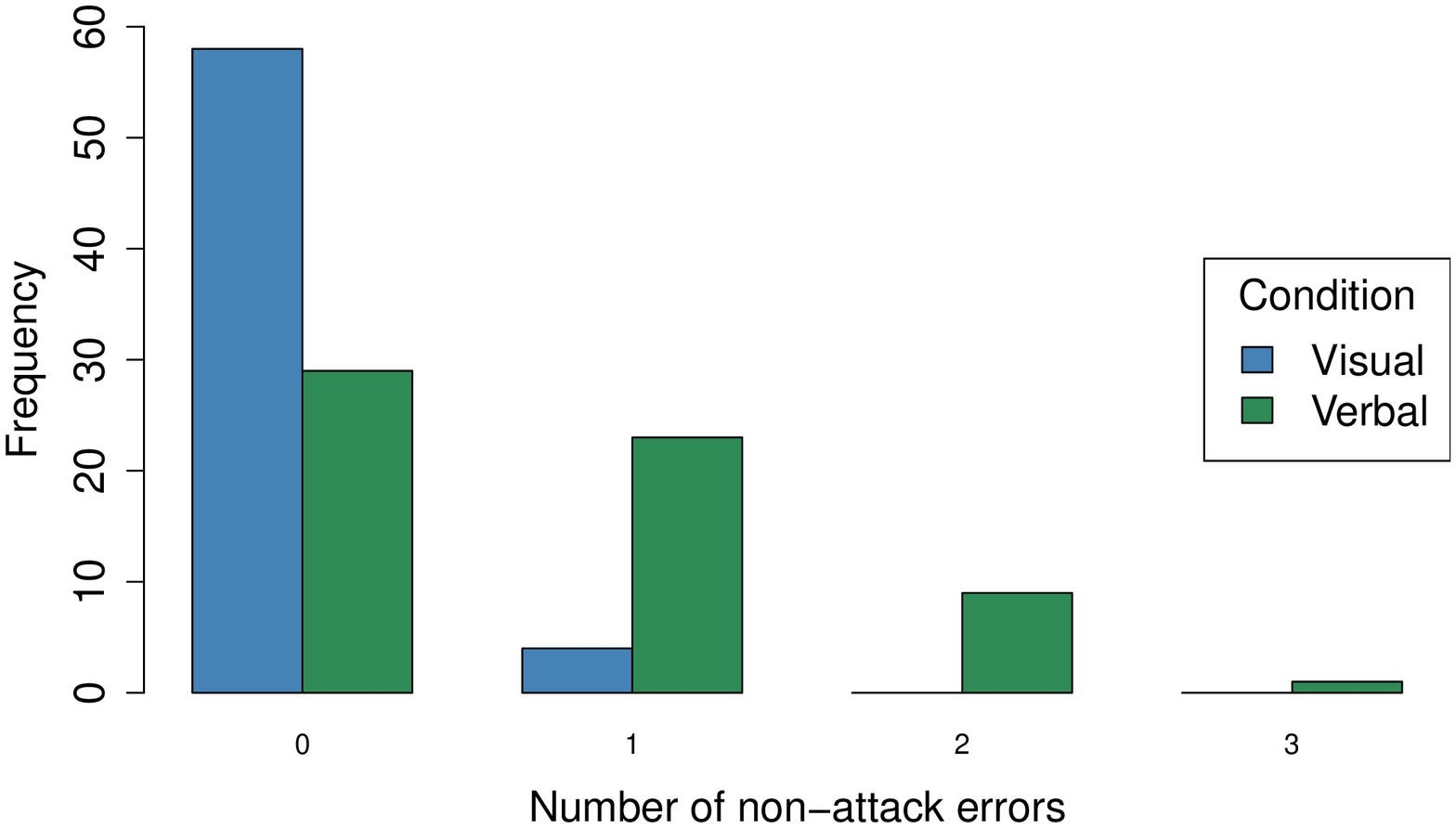}
\label{fig:false_negatives_plot}
\end{minipage}\qquad
\begin{minipage}{.47\textwidth}
\centering
\caption{Number of errors \revision{by each participant} on \revision{2} attack comparisons}
\includegraphics[width=1.0\textwidth]{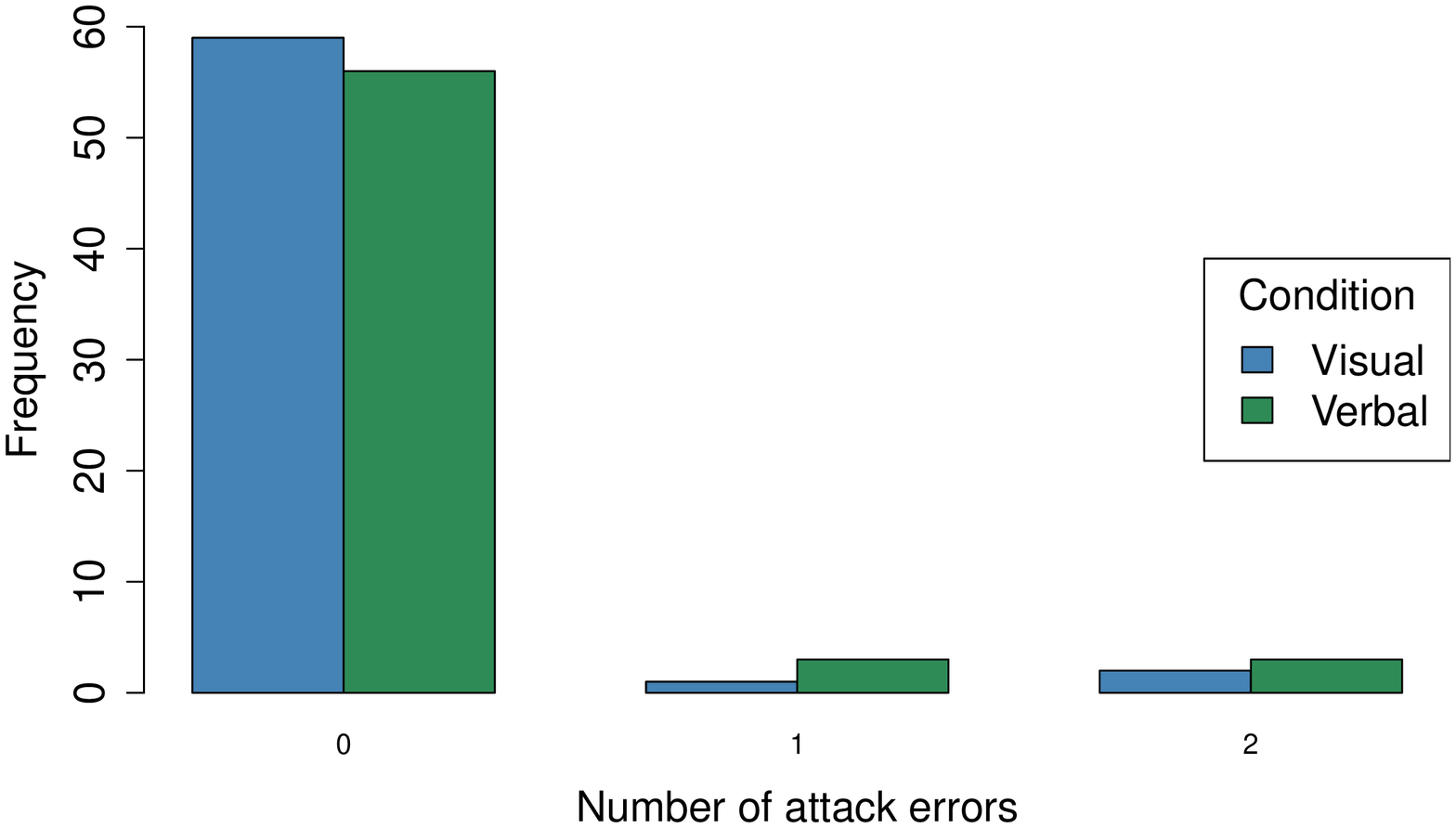}
\label{fig:false_positives_plot}
\end{minipage}
\end{figure}

In general, participants did not make many errors (i.e. identifying a non-attack comparison as an attack or missing an attack comparison).  There were only 2 attack comparisons in each condition, so errors could range from $0$ to $2$. There were 17 non-attack comparisons, so errors could range from 0 to 17. Figs.~\ref{fig:false_negatives_plot} and \ref{fig:false_positives_plot} show the distribution of errors for the non-attack and attack comparisons. There was a difference in errors between the two conditions, with participants making significantly more errors in the verbal non-attack condition than in the visual non-attack condition (see Table~\ref{tab:effectiveness_tests}). Thus $H_2$, that there will be a difference between the errors on the two conditions, was supported.

\begin{table}
\centering 
\caption{Median errors on correct comparisons and SIQR for verbal and visual comparison conditions with Wilcoxon Signed Rank tests of differences between conditions }
\begin{tabular}{|l|c|c|c|c|}
\hline
 & \textbf{Verbal} & \textbf{Visual} & \textbf{Wilcoxon W} & \textbf{p-value} \\ \hline
% \textbf{Attack comparisons}     & 5.49 (0.75)   & 5.50 (1.04)  & 0.22            & 0.83                 \\ \hline
% \textbf{Non-attack comparisons} & 6.15 (0.55)   & 6.52 (1.96)  & 1.20            & 0.23                 \\ \hline
\textbf{Attack comparisons}     & \revision{0 (0.0)}   & \revision{0 (0.0)}  & 1.19            & 0.23                 \\ \hline
\textbf{Non-attack comparisons} & \revision{1 (0.5)}   & \revision{0 (0.0)}  & 4.84            & \textless 0.01       \\ \hline
\end{tabular}
\label{tab:effectiveness_tests}
\end{table}

\subsection{Perceived usability and related concepts}

The ratings on the two items for all six dimensions of perceived usability and related concepts were all highly correlated (Spearman’s $\rho$ %rho
between 0.31 and 0.82, all $p < 0.01$), so median scores were calculated for each dimension and used in subsequent analyses. 
Table~\ref{tab:usability_tests} shows participants’ median ratings for the six dimensions for the visual and verbal conditions. \revision{There was a significant difference on the low mental workload dimension ($p < 0.01$), with the verbal condition perceived to require less mental workload than the visual condition.} There was a strong trend towards a difference on the ease of use dimension ($p = 0.06$), with the verbal condition rated as easier than the visual condition. 
There was also a significant difference on the confidence dimension ($p = 0.02$). Although the median ratings were the same, inspection of the distributions showed that more participants had confidence in the visual condition than the verbal condition. These results show partial support for $H_3$, that there is a difference in the perceived usability of the two conditions, with the verbal condition being perceived as more usable on two out of six dimensions. 
In addition, at the end of the study, participants were asked which comparison mode they would prefer to use, verbal or visual. There was an almost even split between preferences for each system, with 53.2\% choosing verbal and 46.8\% choosing visual.  This was not a significant difference ($\chi^2 = 0.26$, $p = 0.61$).

\begin{table}
\centering
\caption{Median ratings (with SIQR) of the perceived usability dimensions for verbal and visual conditions and Wilcoxon Signed Rank tests of differences between conditions}
\begin{tabular}{|c|c|c|c|c|}
\hline
\textbf{Dimension} & \textbf{Verbal} & \textbf{Visual} & \textbf{Wilcoxon W} & \textbf{p-value} \\ \hline
\textbf{Efficiency}                    & 4.00 (1.00)                                                                                                        & 4.00 (1.00)                                                                                                                & 0.22                               & 0.83                 \\ \hline
\textbf{Ease of use}                    & 4.50 (0.50)                                                                                                        & 4.25 (0.75)                                                                                                                & 1.84                               & 0.06                 \\ \hline
\textbf{Low mental workload}               & 4.00 (0.82)                                                                                                        & 3.00 (1.00)                                                                                                                & 4.21                               & \textless{}0.01      \\ \hline
\textbf{Confidence}                    & 4.50 (0.50)                                                                                                        & 4.50 (0.75)                                                                                                                & 2.39                               & 0.02                 \\ \hline
\textbf{Repeat use}                    & 4.00 (0.75)                                                                                                        & 3.50 (1.00)                                                                                                                & 1.35                               & 0.18                 \\ \hline
\textbf{Trust}                         & 4.00 (0.82)                                                                                                        & 4.00 (1.00)                                                                                                                & 0.76                               & 0.45                 \\ \hline
\end{tabular}
\label{tab:usability_tests}
\end{table}

\subsection{Effect of preferred information style: Verbal versus Visual}

The participants’ scores on the Visual--Verbal subscale of the ILS were skewed towards the visual end of the scale. To create groups of approximately equal size for analysis, participants were divided into three groups: Very Visual (scores 7 to 11, 23 participants); Moderately Visual (scores 1 to 5, 21 participants); and Verbal (scores $-1$ to $-9$, 18 participants). 

There were no significant differences in time to complete correct comparisons in either the verbal or visual conditions between the three groups of participants.  Nor were there any significant differences in the errors made on the attack comparisons.  However, all three groups made significantly more errors in the verbal condition than in the visual condition (Wilcoxon related samples tests, Very Visual: $W = 2.95$, Moderately Visual: $W = 2.88$, Verbal: $W = 2.64$, all $p < 0.01$).  This does not support $H_{4}$,
which predicted verbal users make more errors on the visual condition and visual users make more errors on the verbal condition.

%% file: sections/discussion_conclusions.tex
\section{Discussion and Conclusions}

\revision{This paper reported the results of the first investigation of differences in effectiveness, efficiency and perceived usability between visual and verbal comparisons of word-based key fingerprints.}

\revision{Participants were found to make more non-attack errors when using a verbal comparison mode. One explanation for this result is that it is easier to mishear than misread a word. Without asking for the word to be spelt out, users are unable to check the spelling of any unfamiliar spoken words, and this uncertainty may cause users to reject fingerprints that they would otherwise accept if a visual comparison mode was used. This explanation gains further support since participants perceived that the visual condition provided increased confidence that they were getting the comparisons correct.  In contrast, the verbal condition was perceived to require less mental effort and be easier to use. Since fingerprint comparisons are a secondary task to actual communication, these factors %may be of greater importance to real users, and 
may motivate them to choose a verbal comparison mode even though visual comparisons would provide increased effectiveness and confidence.}

\revision{Even though visual comparisons were shown to be effective and perceived to provide increased usability in two of the six dimensions assessed, practical examples of secure messaging applications largely encourage the use of a verbal comparison mode and tend not to support or encourage visual comparisons. Given these findings, it seems some users would benefit from applications adding increased support for both visual and verbal fingerprint comparisons.}

A surprising result was the lack of effect between comparison mode and Visual--Verbal subscale score. One interpretation is that the main effect of comparison mode dominates, and visual comparisons are significantly more effective \revision{against non-attack errors} for all users. However, care must be taken before reaching this conclusion given the sample's skew towards participants with a visual preference to receive and process information. Further research, that includes a greater proportion of participants with a verbal preference, is required to clarify this.
Another explanation is that the Visual--Verbal subscale does not measure the intended phenomena and an alternative scale may be more appropriate. 7 of the 11 Visual--Verbal subscale questions actually provide 2 visual responses (e.g. written text or diagrams). Future work will attempt to identify a measure of difference between auditory and visual preferences to receive information.

All the fingerprints in this study were based on the Trustwords representation of PEP over PGP. The Trustwords word base contains many unusual and unfamiliar words which may have contributed to the increased number of non-attack errors in the verbal condition. 
% as participants were unfamiliar with their spelling or pronunciation. 
Future research may include fingerprints in other representations (e.g.\ the numeric representation used by Signal/WhatsApp) to determine if the effects observed in this study are specific to the Trustwords representation or fundamental properties of a fingerprint verification.   

A limitation of the study was that each condition included only two attacks. Though there were good reasons for the low attack rate, it made identification of a significant effect between conditions difficult. Furthermore, attacks lacked enough similarity and participants identified them with ease. Future work will include a greater number of attack trials that display greater similarity.

%\revision{A within-participants study with 62 participants investigated differences in %effectiveness, efficiency performance and perceived usability between visual and verbal comparisons of word-based key fingerprints, and the related influence of a participants cognitive learning style.} 

\revision{The answer to which comparison mode is best remains unclear. Visual comparisons were found to be more effective against non-security errors and perceived to provide increased confidence, yet verbal comparisons were perceived to be easier and require less mental effort.}
\revision{Though participants often displayed a preference for a particular comparison mode (based on measures of both performance and perceived usability), this did not correlate with their score on the Visual--Verbal subscale of the ILS.}
\revision{The results show that identification of the optimal comparison mode and the related influence of a user's cognitive learning style on key fingerprint comparisons remain unclear. 
These present complex and interesting research questions that require further investigation.}